\documentclass[a4paper,fleqn]{cas-sc}

\usepackage[numbers,sort&compress]{natbib}
\usepackage[utf8]{inputenc}
\usepackage{booktabs}
\usepackage{amsmath}
\usepackage{xurl}
\usepackage{url}

\IfFileExists{dsfont.sty}{\usepackage{dsfont}}{\providecommand{\mathds}{\mathbb}} 

\RenewDocumentCommand\printorcid{}{}
\ExplSyntaxOn
\cs_set:Npn \__first_footerline: { \mbox{} }
\ExplSyntaxOff

\begin{document}
\let\WriteBookmarks\relax
\def\floatpagepagefraction{1}
\def\textpagefraction{.001}

\shorttitle{EV charging under power tariffs and spot prices in Swedish LV grids}
\shortauthors{E. Perotti et al.}

\title[mode=title]{Electric vehicle charging under household power tariffs and spot prices: Impacts on residential low-voltage grids in Sweden}

\author{Elisabetta Perotti}
\cormark[1]
\fnmark[1]
\ead{elisabetta.perotti@chalmers.se}

\author{Therese Lundblad}
\fnmark[1]

\author{Maria Taljegard}
\author{Niclas Mattsson}
\author{Filippo {Da~Corte~Vecchino}}
\author{Filip Johnsson}

\affiliation{organization={Department of Environmental and Energy Sciences, Chalmers University of Technology},
            city={Gothenburg},
            postcode={412 96},
            country={Sweden}}

\cortext[1]{Corresponding author}
\fntext[1]{E. Perotti and T. Lundblad contributed equally to this work.}

\begin{abstract}
Home charging of electric vehicles (EVs) is increasingly steered by price signals, such as electricity spot prices and network tariffs on the household peak power. This paper assesses how cost-minimizing charging under spot prices, with and without a power tariff on the monthly household peak, affects thermal and voltage violations in residential low-voltage grids across Sweden. Results are compared with logged real-world charging for EV fleet shares up to 100\%. The optimized charging profiles, computed jointly with measured household loads, are allocated to synthetic grids covering 5.1 million customers, calibrated against distribution system operator data. Spot-price charging without a power tariff more than doubles the peak power of four in five transformers at 100\% EVs and overloads the transformer of 42\% of customers, against 0.5\% with logged charging. The power tariff reduces this share to 6.9\%, but it still requires new transformer capacity of about 7\% of the installed capacity, because households responding to the same prices continue to charge simultaneously even when their individual peaks are lower. The power tariff brings transformer peaks below those of logged charging only where very few customers share a transformer, so results from single grids cannot be generalized. If cost-minimizing charging becomes widespread, tariffs on the combined peak of a group of households may be needed to keep charging within the capacity of existing grids.
\end{abstract}

\begin{keywords}
Demand charges \sep distribution grids \sep electric vehicles \sep home charging \sep low-voltage grids \sep power tariffs \sep synthetic grids
\end{keywords}

\maketitle

\section{Introduction}

Electrification of road transport is a key measure for reaching climate targets, requiring a substantial increase in the share of electric vehicles (EVs) in the passenger vehicle fleet worldwide~\cite{EuropeanEnvironmentAgency}. Since most EVs are charged at home, most of the additional electricity demand will be connected to the low-voltage (LV) distribution grid, which was dimensioned for residential loads without EVs~\cite{Nour2020, Muratori2018}. Whether the existing LV grids can host this demand without exceeding their operational limits depends not only on the number of EVs, but also on when they are charged.

The timing of home charging is increasingly shaped by price signals. The price signal used to control EV charging can be based on any of the components of the electricity price or combinations thereof. The retail price of electricity can be split into the wholesale electricity price, network tariffs, and taxes and levies~\cite{ACER2025}. For Sweden, the wholesale price is cleared on the Nord Pool spot market and is often referred to as the spot price. It reflects the balance between supply and demand in the whole electricity bidding zone and carries no information about the loading of the local grid. Network tariffs are paid to the distribution system operator (DSO), and reflect the costs for operating, maintaining, and expanding the grid infrastructure. European DSOs are moving towards network tariffs based on peak power~\cite{ACER2025}. In Sweden, many DSOs have introduced such power tariffs for residential customers, with large differences in design: some charge the monthly peak at all hours, while others only consider peaks during daytime hours on weekdays or during the winter months~\cite{Lundblad2026preprint}. The design of these tariffs is currently under debate: in 2026, the Swedish government stopped the requirement for DSOs to introduce power tariffs and tasked the Swedish Energy Markets Inspectorate (Ei) with proposing a new design~\cite{TheGovernmentOfficesofSweden2026}. Whether a power tariff relieves the grid when households also respond to the spot price is therefore a question of direct relevance for regulators and DSOs. 

\subsection{Literature Review}

The impact of EV charging on distribution grids has been studied extensively, typically by estimating the hosting capacity, i.e., the amount of EVs that a grid can accommodate before its operational limits are exceeded~\cite{ZainulAbideen2020, Carmelito2023, Umoh2023, Mulenga2020}. A consistent finding is that uncontrolled charging upon arrival at home coincides with the evening peak of household demand~\cite{Nour2020, Muratori2018}, while charging that responds to a common price signal can create new and higher peaks, since all EVs shift their charging to the same low-price hours~\cite{Schachler2021, Veldman2015, Sandstrm2023}. Schachler et al.~\cite{Schachler2021} modeled how EV charging based on a single electricity price signal affects the medium-voltage (MV) grids in Germany, with comparison to a reference case that used an uncontrolled charging strategy. They considered a scenario for Year 2030 with 10 million EVs, and showed that the impacts of different charging strategies differed between different regions of Germany. In urban areas, EV charging based on an electricity price signal led to a significant increase in the number of grid issues, as compared with the uncontrolled charging strategy. In non-urban areas, there were few instances in which the MV grid capacity was exceeded. Veldman and Verzijlbergh~\cite{Veldman2015} found that, for Dutch distribution networks,  cost-minimized charging caused high load peaks requiring grid reinforcement, whereas charging that minimized household peak demand substantially reduced reinforcement needs. However, the latter strategy did not consider charging costs and their variation over time. Zaferanlouei et al.~\cite{Zaferanlouei2022} modeled a Norwegian grid with 856 customers and found that direct charging caused thermal violations beyond EV fleet shares of 18--27\% and cost-minimized charging beyond 36\%, whereas only charging scheduled with full knowledge of the grid constraints could host 100\% EVs. In a case study of a Swedish residential area, Sandstr\"{o}m et al.~\cite{Sandstrm2023} found that the hosting capacity depended strongly on the charging strategy and pointed to power tariffs as a likely driver of future charging behavior. They suggested that future work should investigate EV charging that responds to electricity price signals while limiting the household peak power. Gan et al.~\cite{Gan2013} proposed a decentralized protocol in which the utility company repeatedly broadcasts a control signal based on grid loading. However, this approach requires continuous communication between the utility and individual EVs, potentially creating barriers to implementation.

A separate body of literature has studied how network tariffs affect household demand. Power-based tariffs have been shown to reduce household peak demand and to allocate network costs more in line with the costs that customers cause~\cite{Turk2024, Saele2017, Azuatalam2017}. Velkovski et al.~\cite{Velkovski2024} found that a power component in the tariff reduced the peak loads of an energy community, and Abdelmotteleb et al.~\cite{Abdelmotteleb2018} showed that a tariff charging customers for their contributions to peak hours improved economic efficiency in a simple two-bus network. Daneshzand et al.~\cite{Daneshzand2023} found that smart charging under British retail tariffs could increase local grid stress, which active capacity management could alleviate. El Gohary et al.~\cite{ElGohary2023} pointed out that individual customers' peak demand often does not coincide with the system peak, so that power tariffs may steer customers towards behavior of limited value for the grid. Panda et al.~\cite{Panda2024} studied how segmented network tariffs influence the aggregate peak demand of EV charging, and Xiao et al.~\cite{Xiao2025} designed a time-of-use tariff with a demand charge for residential charging posts. Lundblad et al.~\cite{Lundblad2026preprint} developed a model to optimize the home charging of individual EVs together with a measured household load under the spot price and a network tariff, and used it to compare power tariff designs applied in Sweden. For 188 households, a tariff on the monthly peak reduced their aggregated peak by 33\% compared with charging against the spot price only, whereas a tariff limited to daytime hours mainly shifted the peaks outside the tariff hours.
 
These two bodies of literature have rarely been combined, and not at national scale. Grid impact studies have represented price-responsive charging mostly through the spot price alone, or through grid-aware scheduling that assumes the grid constraints are known to a central actor~\cite{Zaferanlouei2022, Gan2013}. Tariff studies, in turn, have mostly evaluated peak demand at the household, community or fleet level~\cite{Turk2024, Velkovski2024, Panda2024, Lundblad2026preprint}, and the few that assess the limits of LV grids consider a small number of distribution feeders~\cite{Jung2023}. The combination of a spot price and a power tariff, which is the situation that some Swedish households with EVs now face and that was explicitly called for in~\cite{Sandstrm2023}, has therefore not been evaluated against LV grid constraints across different regions. 
 
\subsection{Statement of Contributions}
 
This paper couples the EV charging model MANORS~\cite{Lundblad2026preprint} with the synthetic LV grid model REGAL~\cite{Lundblad2024a}, which represents all residential LV grids in Sweden. The contributions of this paper are as follows:
\begin{itemize}
    \item A national-scale assessment of how a power tariff affects thermal and voltage violations and transformer reinforcement needs in residential LV grids, covering all populated $1\times1$~km$^2$ grid cells in Sweden, for EV fleet shares up to 100\% of the privately owned passenger car fleet.
    \item A comparison of three charging cases for the same households and EVs: logged real-world charging and cost-minimizing charging under the spot price with and without a power tariff on the monthly household peak.
    \item An analysis of how the grid impacts of each charging case vary with population density and the number of customers per transformer.
\end{itemize}
To the authors’ knowledge, this is the first work to evaluate how cost-minimizing EV charging under both a spot price and a household power tariff affects violations in the residential LV grids of an entire country.
 
\subsection{Outline}
 
The remainder of this paper is organized as follows. Section~\ref{sec:methods} describes the modeling framework: the synthetic LV grid model, the EV charging model and the charging cases, the representation of houses and apartment buildings, and the evaluation of violations. Section~\ref{sec:results} presents the increase in peak power, violations at increasing EV fleet shares, where and when they occur, and the need for transformer reinforcement. Section~\ref{sec:discussion} discusses the results, their implications for DSOs and tariff design, and the limitations of the study. Section~\ref{sec:conclusions} concludes the paper.
 
\begin{figure*}[pos=!tbp] \centering \includegraphics[width=\textwidth]{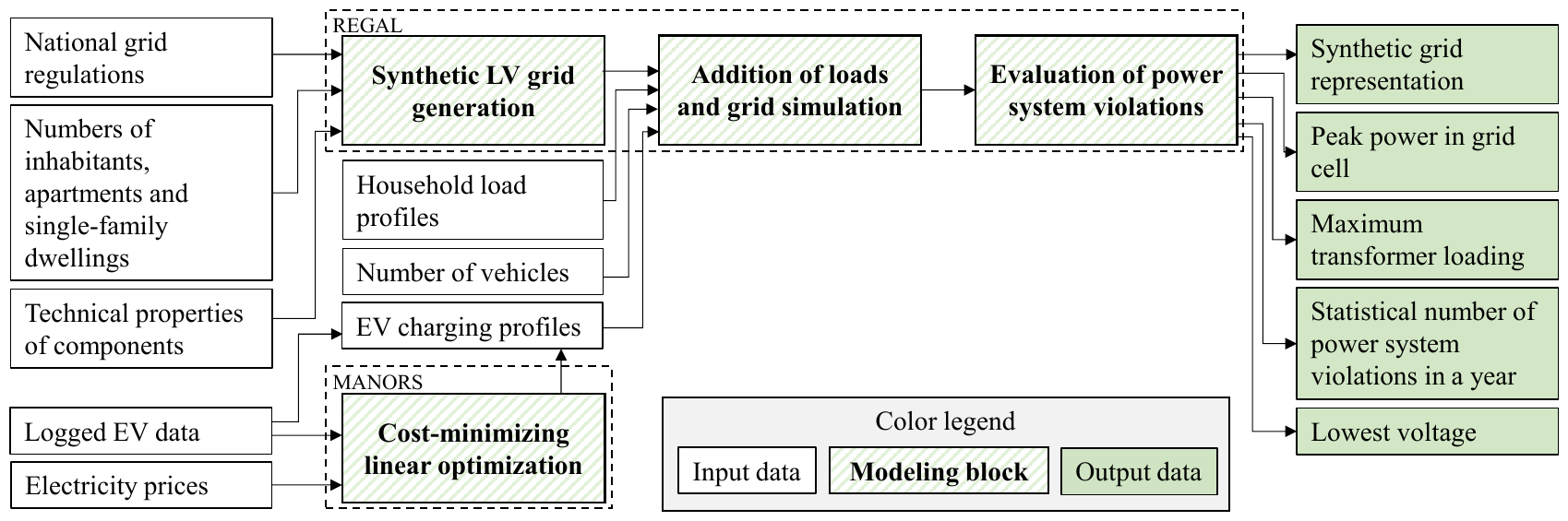} \caption{The EV charging model MANORS optimizes the home charging of each house with one or two EVs; REGAL generates the synthetic LV grid of each cell, allocates loads to the feeder segments, and evaluates violations.} \label{fig:framework} \end{figure*}
 
\section{Method}\label{sec:methods}
 
The study couples two models as shown in Fig.~\ref{fig:framework}. The EV charging model MANORS~\cite{Lundblad2026preprint} generates the load profile of a house including the cost-optimized home charging of its EVs. The synthetic LV grid model REGAL~\cite{Lundblad2024a} places these profiles, together with apartment buildings, on synthetic LV grids covering all residential areas in Sweden and evaluates thermal and voltage violations over one year at 15-min resolution. Both models are open source~\cite{Lundblad2024b, ChalmersEnergySystems2026}.
 
\subsection{Synthetic low-voltage grid model}
 
Detailed data on LV grids are not publicly available in Sweden, where more than 150 DSOs operate local grids. REGAL therefore generates synthetic grids from open demographic and geospatial data and established grid design principles~\cite{Lundblad2024a}. Sweden is divided into 106,777 populated cells of $1\times1$~km$^2$. For each cell, the numbers of single-family houses and apartments are derived from Statistics Sweden (SCB) data~\cite{StatisticsSweden2025a}. Each house and each apartment is one customer, i.e., one household with its own metered connection. From the estimated residential demand, REGAL dimensions the distribution transformers and the feeder cables, meeting Swedish requirements on voltage quality, tripping criteria, and thermal limits of components (Appendix~\ref{app:design}). The design rules are calibrated and validated against DSO data (Section~\ref{sec:validation}). The analysis covers the 400~V network supplying 5.1 million residential customers. Commercial and public customers, such as shops, schools, and offices, are neither simulated nor included when the grid is dimensioned. 
 
Each cell is represented by one representative LV network, i.e., one transformer with its feeders and customers, which is replicated $N_c$ times to serve all customers of cell $c$. Voltage and cable loading are evaluated along the longest feeder, which is divided into segments with its customers distributed evenly along it, and the voltage is evaluated at the furthest customer. The remaining customers of the transformer are connected at the transformer, so that they contribute to the transformer loading and the voltage drop over the transformer. The accuracy of the synthetic grids is high at an aggregated level but lower for individual cells, and results are therefore reported mainly at aggregated levels (Section~\ref{sec:metrics}). The grid design is the same in all charging cases and EV fleet shares; it is not reinforced when EVs are added.
 
\subsubsection{Calibration and validation against grid data}\label{sec:validation}
The grid design is calibrated on DSO data covering about 10\% of the populated cells in Sweden, spread over seven regions, including the number and installed capacity of transformers per cell, the measured loading of the transformers and the length of LV cables per type. Consistent with the residential-only design, the calibration uses mainly residential cells, in which residential customers account for most of the metered energy (more than 80\% for house areas and more than 60\% for apartment areas). The design has few free parameters: the number of customers per transformer, three transformer margins (for small networks, large networks, and apartment-dominated areas) and one cable margin (Appendix~\ref{app:design}).

The robustness of the calibration is tested by leaving out each of the seven regions in turn, refitting the parameters on the remaining regions, and comparing the result with the region left out. The refitted parameters differ from the ones used by at most 0.2 (Table~\ref{tab:loro}), so the design does not depend on the data of any single region.

In the mainly residential cells of the calibration data, the modeled number of transformers matches the real one exactly in 63\% of the cells and within $\pm1$ in 92\%; in total it is 17\% lower, mainly in small rural cells, where the real grids also supply holiday homes that are not in the model. The installed capacity matches the real one in the median cell, but in total it is 26\% lower, mainly due to cells with more than 100 customers where the real transformers are larger, likely including some non-residential load.
 
\subsubsection{Power flow approximation}
 
Voltages and loadings are calculated with a linearized radial power flow~\cite{Baran1989}, which makes year-long simulations of all cells feasible. The voltage at the furthest customer of the representative LV network in timestep $t$ is
\begin{equation}
U_{t} = U_{\mathrm{N}} - \frac{R_{\mathrm{T}} P_{0,t} + X_{\mathrm{T}} Q_{0,t}}{U_{\mathrm{N}}} - \sum_{j=1}^{J} \frac{R_{j} P_{j,t} + X_{j} Q_{j,t}}{U_{\mathrm{N}}},
\label{eq:voltage}
\end{equation}
where $U_{\mathrm{N}}=400$~V, $R_{\mathrm{T}}$ and $X_{\mathrm{T}}$ are the transformer resistance and reactance, $R_j$ and $X_j$ are those of feeder segment $j$, and $P_{j,t}$ and $Q_{j,t}$ are the active and reactive power flowing through segment $j$, i.e., the sum of the loads of all customers downstream of it ($P_{0,t}$, $Q_{0,t}$: all customers of the transformer). The no-load voltage on the low-voltage side of the transformer is fixed at $U_{\mathrm{N}} = 400$~V in every timestep, so voltage variations in the medium-voltage grid and the transformer tap setting are not represented.
 
The active power is the load of the customers, including EV charging. The reactive power is attributed only to the household and common-area loads, which are assumed to have a power factor of $\cos\varphi = 0.9$. EV chargers are assumed to operate at unity power factor. 
 
\subsection{Residential loads}\label{sec:loads}

Residential demand is represented by measured load profiles with 15-min resolution, collected between April 2024 and March 2025 by Mölndal Energi DSO. After cleaning, 3,262 profiles of single-family houses and 7,645 profiles of apartments are available, covering electric and non-electric heating. 
 
\paragraph{Houses} Each house has zero, one, or two EVs. The houses are represented by 1,000 of the house profiles, for which the charging is optimized with MANORS. Since an LV network contains at most 108 houses in REGAL, 1,000 profiles give enough variety while significantly reducing computational time. Of these houses, 80\% are electrically heated; they use on average 14.6~MWh per year, against 6.4~MWh for non-electrically heated houses. The fuse size of each house is taken from the metadata of its profile: 54\% of the houses have 20~A, 33\% 25~A, 11\% 16~A, and 2\% 35~A. A house without EVs is represented by its household load, a house with EVs is represented by the load profile from MANORS, in which the EV charging is optimized together with the household load. In REGAL, a whole number of houses is drawn to match the expected number of houses in the cell. Each house is then assigned to the class with zero, one, or two EVs at random, with the class shares as probabilities. If a class has at most 10 houses in a segment, these houses are drawn as individual households, and their profiles are summed directly. For larger groups, precomputed aggregates of increasing size have been used and rescaled to the exact number.
 
\paragraph{Apartment buildings} The apartments use on average 2.35~MWh per year (2.05~MWh without and 8.2~MWh with electric heating). A pool of buildings is built from the apartment profiles where electrically heated profiles make up about 5\% of the total. In the pool, each building is the sum of $N$ different apartment profiles and $\mathrm{round}(rN)$ EV charging profiles, for building sizes $N \in \{4, 6, 9, 12, 18, 25, 35, 50, 75, 100\}$ apartments and EV rates $r \in \{0, 0.25, \dots, 1.75\}$ EVs per apartment, with 100 buildings per size class. A building unit has the same apartments in every EV-rate class, and the EVs of a lower rate are a subset of those of a higher rate, so that a higher EV fleet share gives the same building with more EVs. In this work, EV charging in apartment buildings is not optimized. The real-world logged charging profiles are therefore used in all charging cases. For each segment of a cell, buildings are drawn until they match the segment number of apartments. The size class of each building is drawn from a lognormal distribution whose mean equals the cell average number of apartments per building, which was derived from SCB data~\cite{StatisticsSweden2025a}. 
 
\paragraph{Common areas} Each apartment building also has a load for its common areas, with an annual energy of 1,200~kWh per apartment. This value is calibrated so that the modeled consumption of apartment areas matches the residential and collectively metered consumption in DSO data (Section~\ref{sec:validation}). The load is modeled using a synthetic 15-minute profile that combines ventilation with reduced nighttime operation, lighting demand that follows the seasonal daylight cycle, and elevator usage characterized by morning and evening peaks.
 
\paragraph{Consumption factor} Real neighborhoods differ significantly in their consumption, for example through the type of heating and the size of the houses, whereas houses and buildings drawn from the measured profiles make every area consume like an average Swedish area. The household load of the houses and that of the apartment buildings in cell $c$ is therefore multiplied by a cell consumption factor $f_c = \exp(\sigma z_c - \sigma^2/2)$, a lognormal factor with mean 1, where $z_c$ is a standard normal variable common to the houses and buildings of the cell. EV charging is not scaled. The standard deviations, $\sigma = 0.3$ for houses and $\sigma = 0.2$ for apartment buildings, are calibrated on the spread of the measured loading of transformers in mainly residential areas (Section~\ref{sec:validation}). Since the consumption factor of a particular cell is not known, $z_c$ is treated as part of the uncertainty about the cell, and it enters the simulations in two ways depending on the result considered (Section~\ref{sec:metrics}).
 
\subsection{EV data, charging model and cases}\label{sec:ev}
 
\paragraph{EV data} EV driving and charging are based on GPS logging of 139 EVs in Sweden selected from a larger dataset described in detail by Kobayashi et al.~\cite{Kobayashi2025}. The selected EVs drive on average 15,700~km and charge 2,250~kWh at home per year. EVs charge at home on average at 4.7~kW, with an average individual peak of 7.1~kW. The peak of the 139 summed profiles is 1.17~kW per EV. The data give, per 15-min timestep, the energy used for driving, the energy charged, and whether the vehicle is at home. To avoid that a limited set of cars repeats across many households, each car time series is shifted by a whole number of weeks between $-3$ and $+3$, giving 973 EV profiles. The same EV set is used for houses and apartment buildings: only 44 of the logged owners live in apartments, too few to form a separate set, so the charging of apartment dwellers is represented by the charging behavior of all logged owners. 37\% of the EV owners have an hourly electricity contract: 44 of the 94 house owners, but only 6 of the 44 apartment owners (the dwelling type of one owner and the contract type of another are unknown). Owners with an hourly electricity contract charge 59\% of their home energy between 22:00 and 06:00, against 40\% for owners without one, and concentrate it between 00:00 and 03:00.

The selected EVs have at least one year of data without long gaps and total charging consistent with the driving energy; missing values (5\% of the timesteps on average) are filled with data from the same EV one to three weeks earlier or later.
 
\paragraph{EV charging model} MANORS~\cite{Lundblad2026preprint} is run with the house profiles of this study and with one or two EVs per house. It minimizes the annual electricity cost of each household: the purchase of electricity at the spot price, public charging away from home at a fixed price of 0.56~€/kWh~\cite{EuropeanCommission-EuropeanAlternativeFuelsObservatory2024}, and, where applicable, a monthly power tariff. The household load is fixed; the home charging of the EVs is optimized subject to the energy balance of each battery, the times the EV is parked at home, one home charger per house (shared by its EVs when there are two), a charging efficiency of 95\%, and the household's main fuse, which limits the sum of household load and charging to at most 110\% of the fuse rating. The maximum charging power is set to 11 kW, the most common on-board charger capacity for battery EVs registered in Sweden since 2021~\cite{MobilitySwedenNyregistreringar}. The logged EVs charge on average at lower power; part of the difference between the logged and the optimized cases therefore reflects the higher charging power adopted in the latter.
The driving demand is always met, if needed through public charging, so no demand is left unserved. The model is a linear program solved for one year with perfect foresight of prices, driving, and household demand. This does not predict real behavior; it shows the incentives each tariff design creates, while the logged charging provides a reference for current behavior. 
 
\paragraph{Time alignment} All profiles are represented on a common year of 35,040 timesteps. The EV time series were logged in different periods and are stitched so that each starts on 1 January~\cite{Lundblad2026preprint}. The household and EV profiles are aligned so that their weekdays coincide, but calendar dates are not matched exactly between household load, EV use, and spot prices. Since the analysis concerns the statistical coincidence of loads over a year rather than specific dates, this is not expected to affect the comparison between charging cases.
 
\paragraph{Charging cases} Three cases are compared:
\begin{itemize}
    \item \emph{Logged}: logged home charging, without optimization.
    \item \emph{No tariff}: charging cost-minimized against the spot price only.
    \item \emph{Power tariff}: charging cost-minimized against the spot price and a power tariff on the monthly household peak.
\end{itemize}
Both optimized cases apply to individual household connections and are therefore applied to houses only. Differences between the charging cases thus arise entirely from houses, and in areas dominated by apartment buildings the cases are expected to differ little.
The power tariff is 7.4~€/kW, the average of 24 Swedish DSOs~\cite{Lundblad2026preprint}, charged on the household's highest 15-min average power per month. Spot prices are the hourly Nord Pool prices of 2024~\cite{NordPoolSpot2025} in each of the four Swedish bidding zones (SE1--SE4), and each cell uses the profiles of its bidding zone. The effect of the price year on charging is analyzed in~\cite{Lundblad2026preprint}. The logged charging case does not depend on the bidding zone.
 
\paragraph{Vehicle allocation}
 
EVs are introduced where privately owned passenger cars are currently registered. Car ownership per house and per apartment is derived per cell from SCB data~\cite{StatisticsSweden2025a}. Scenarios are defined by the EV fleet share, i.e., the share of these cars that are EVs, of 25, 50, 75 and 100\%. At 100\%, the model contains 4.23 million EVs. Houses can have at most two EVs: with $c_{\mathrm{H}}$ EVs per house on average, the probability that a house has one EV is $c_{\mathrm{H}}$ and that it has two EVs is zero if $c_{\mathrm{H}} \leq 1$; if $c_{\mathrm{H}} > 1$, every house has at least one EV and the probability of two is $c_{\mathrm{H}}-1$. Apartment buildings have at most 1.75 EVs per apartment. Numbers of houses and EVs are drawn as whole numbers whose expected values match the data in each cell. 
 
\subsection{Evaluation of violations}\label{sec:metrics}
 
\paragraph{Criteria} Three types of violations are evaluated in every timestep: \emph{undervoltage}, when the voltage at the furthest customer is more than 5\% below nominal ($U_t < 380$~V); \emph{transformer overload}, when the transformer loading exceeds its rated power; and \emph{cable overload}, when the current in any feeder segment exceeds the cable rating. The 5\% limit is stricter than the $\pm10\%$ allowed by the Swedish regulation EIFS~2023:3~\cite{EIFS2023_3}. The criterion is also stricter than the standards in that every timestep outside the limit counts as a violation, whereas EN~50160~\cite{EN50160} requires 95\% of the 10-min values of a week to lie within the range.
 
\paragraph{Household draws} Which households, EVs, and buildings are connected where in a real grid is unknown. Each cell is therefore simulated many times, over the whole year, each time with a new random combination of households and buildings, here referred to as a \emph{draw}. The same households, buildings, and consumption factors are drawn in every charging case and EV fleet share: a house keeps its household demand when its number of EVs increases with the fleet share. Differences between cases and fleet shares are thus not blurred by different random compositions. 

Let $v_{c,x,t,s}\in\{0,1\}$ indicate a violation of type $x$ in cell $c$, timestep $t$, and draw $s$, where several overloaded segments in the same timestep count once. The number of violating timesteps in draw $s$ is
\begin{equation}
n_{c,x,s} = \sum_{t=1}^{T} v_{c,x,t,s}.
\end{equation}

\paragraph{Two sets of runs} The consumption factor enters the draws in two ways, suited to different kinds of results. In the \emph{random-level runs}, $z_c$ is drawn anew in every draw, so each draw combines a set of houses, buildings, and EVs with a new random consumption factor of the cell. Since the factor has a mean of 1, it does not change the expected household consumption of a cell, only its spread between cells and years. These runs use $S = 100$ draws per cell at all EV fleet shares and in all charging cases. They are used for all results that are means, sums, or medians over draws: the shares and numbers of customers and LV networks with violations, where and when violations occur, the number of violating intervals, and the increase in peak power. Results differ by less than 1.2\% between random seeds and by less than 2\% between 100 and 1,000 draws per cell, well below the differences between the studied cases (Appendix~\ref{app:convergence}). 

In the \emph{fixed-level runs}, each cell is instead simulated at a set of fixed consumption factors. The range of $z_c$ is divided into 9 intervals; each interval forms one level, with a weight equal to its probability and a factor equal to the mean of the lognormal factor within the interval, so that the weighted mean of the factor is exactly 1. The intervals are narrower in the upper tail, where the results are most sensitive to the consumption factor (Appendix~\ref{app:levels}). Each cell is simulated with 112 draws at each level ($S = 1008$), every per-cell result is computed separately for each level, and the levels are then combined with their weights. These runs are used for results defined by a threshold or a percentile per cell: the peak loading and lowest voltage, the grouping of LV networks by how often they experience violations, the longest run of consecutive violating intervals, and the transformer reinforcement. If $z_c$ were drawn anew in every draw, as in the random-level runs, the unfavorable draw of most cells would come from a high consumption factor, and cells would be judged as if they were high-consumption cells. With fixed levels, the unfavorable draw reflects the variation in composition at the cell’s own consumption factor, and the probability of each level enters through its weight. Because the factor is constant within each level, the fixed-level runs understate tail magnitudes such as the number of violating intervals, which are therefore taken from the random-level runs. The fixed-level runs are made at 0\% and 100\% EVs only.
 
\paragraph{Violating intervals} The loads are 15-min averages of metered energy, so a violating timestep means that the average over the interval exceeded the limit, not that the violation lasted 15~min; short peaks within intervals are not resolved. Violations are therefore reported as numbers of violating 15-min intervals rather than as durations in hours.
 
\paragraph{Metrics per LV network} Three aspects are reported separately: how likely, how frequent, and how large violations are. All measures are first computed within each draw, i.e., within one simulated year of one possible composition, and then summarized over the draws.
\begin{itemize}
    \item \emph{Probability of violation}: the share of draws with at least one violating interval,
    \begin{equation}
    p_{c,x} = \frac{1}{S}\sum_{s=1}^{S} \mathds{1}\left[n_{c,x,s} \geq 1\right].
    \end{equation}
    Every exceedance counts as a violation. $p_{c,x}$ estimates the probability that the LV network, whose actual composition is unknown, violates its limits at least once in a year. In the fixed-level runs, $p_{c,x,b}$ is computed in the same way at each level $b$, with $w_b$ the probability of the level.
    \item \emph{Frequency and duration}: the expected number of violating intervals per year, $\bar{n}_{c,x} = \frac{1}{S}\sum_{s} n_{c,x,s}$, and the longest run of consecutive violating intervals in the median violating draw. Note that $\bar n_{c,x}$ alone does not distinguish rare years with many violating intervals from frequent years with few, but together with $p_{c,x}$ it does, since their ratio $\bar n_{c,x}/p_{c,x}$ is the mean number of violating intervals in a violating year.
    \item \emph{Magnitude}: the peak transformer loading relative to the rated power and the lowest voltage at the furthest customer are recorded in each draw. They are reported for the median draw and for an unfavorable draw, i.e., the 95th percentile of the peak loading and the 5th percentile of the voltage over the draws of a cell.
\end{itemize}
 
\paragraph{Shares of customers and LV networks} REGAL simulates one representative LV network per cell, while cell $c$ contains $N_c$ of them. To avoid assumptions about the dependence between transformers in the same cell, each LV network is counted separately, with the violation probability $p_{c,x}$ of the representative network. The expected share of customers served by an LV network with violations is
\begin{equation}
\Gamma_x = \frac{\sum_c C_c\, p_{c,x}}{\sum_c C_c},
\end{equation}
where $C_c$ is the number of customers, i.e., houses and apartments, in cell $c$. The expected share of LV networks with violations is obtained in the same way, with the number of LV networks $N_c$ as weight. Since undervoltage is evaluated at the furthest customer, the customer-weighted share of undervoltage is the share of customers who live in an LV network where the furthest customer has a voltage below 380~V in at least one 15-min interval of a year. Most customers in such a network are closer to the transformer and see a smaller voltage drop. Results are given nationally and per area type: rural (up to 200 inhabitants per km$^2$), town (200--1,000) and city (above 1,000), as classified in REGAL~\cite{Lundblad2024a}. LV networks are also grouped by their number of customers (1--5, 6--20, 21--50, 51--150 and more than 150), since the size of an LV network determines how strongly the loads of its households coincide. 

\paragraph{Increase in peak power} For each draw, the annual peak load of the transformer in the representative LV network is divided by the annual peak of the same households and buildings without their EV charging. The increase in peak power is taken as the median of this ratio over the draws. Because the numerator and denominator of each ratio come from the same draw, with the same composition and consumption factor, the consumption factor largely cancels, and the random-level runs can be used. Medians and percentiles of the increase over groups of LV networks are weighted by the number of LV networks $N_c$.
 
\subsection{Grid reinforcement}\label{sec:reinforcement}
 
The transformer capacity required for an EV fleet share of 100\% is estimated following the approach in~\cite{Lundblad2026thesis}. The design load of the transformer of an LV network is the 95th percentile of its annual peak load over the draws at a given level of the fixed-level runs. An LV network is reinforced if the design load exceeds the rated transformer capacity by more than 10\%, and its transformer capacity is then increased so that the design load is at most 60\% of the new total transformer capacity of the LV network. Following practices described by Swedish DSOs~\cite{Lundblad2026thesis}, the existing transformer is replaced by the smallest standard size that meets this target if it is the only transformer in the cell and is rated below 800~kVA; otherwise, transformers are added, using the smallest standard size that covers the missing capacity. New transformers have standard sizes from 100 to 1,000~kVA, and from 50~kVA in cells whose existing transformer is smaller than 100~kVA. The result per LV network is multiplied by the number of LV networks $N_c$ in the cell. Only transformers are reinforced; cable reinforcement and measures against undervoltage at the feeder end are not included.

\section{Results}\label{sec:results}
 
\subsection{Increase in peak power}\label{sec:res_peak}

\begin{figure}
\centering
\includegraphics[width=\textwidth]{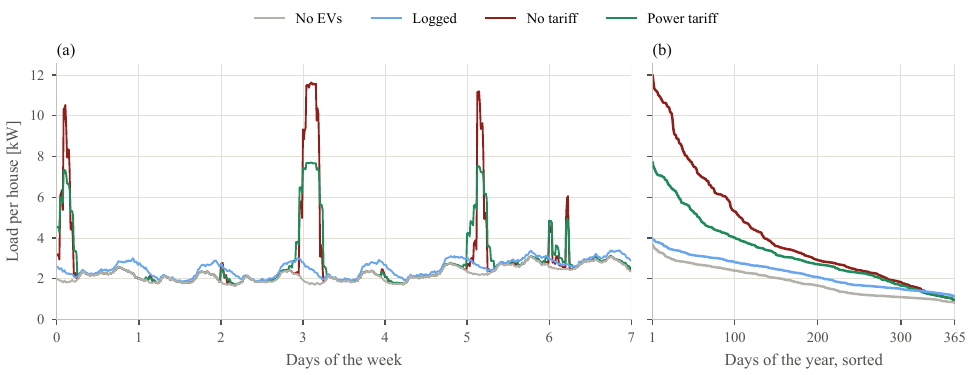}
\caption{Average of 1,000 house load profiles with one EV each, per charging case and without EVs: (a) week around the annual peak of the power tariff case, (b) daily peaks sorted from highest to lowest.}
\label{fig:load_curve}
\end{figure}
 
With logged charging, the 4.23 million EVs draw 10.0~TWh per year from the grid for home charging, about 2,400~kWh per EV including charging losses (95\% efficiency). In both optimized cases, home charging rises to 10.9~TWh (+9\%), since charging at home is cheaper than at public chargers. 

To illustrate the optimized charging patterns with an example, Fig.~\ref{fig:load_curve}(a) shows the average load profile of 1,000 houses each with one EV, over the week that contains the annual peak of the power tariff case. The EVs do not charge every night, but on the cheapest nights they charge together. Fig.~\ref{fig:load_curve}(b) shows the daily peaks sorted over the year. The daily peak exceeds the no-EV annual peak on 13 days with logged charging, 147 days with \emph{No tariff}, and 120 days with \emph{Power tariff}.

\begin{figure*}[pos=!tbp]
\centering
\includegraphics[width=\textwidth]{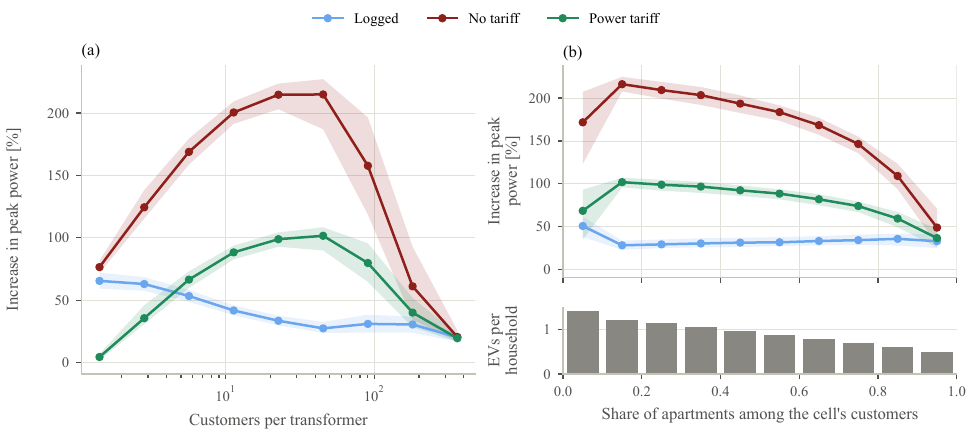}
\caption{Increase in peak power at 100\% EVs relative to the same households without EV charging: (a) as a function of customers per transformer and (b) as a function of the share of apartments among the customers of the cell, with the mean number of EVs per household below. LV networks are grouped by the number of customers per transformer and by the share of apartments. Lines: median of the increase in each group weighted by the number of LV networks; bands: 25th–75th percentile.}
\label{fig:increase}
\end{figure*}

Fig.~\ref{fig:increase}(a) shows the increase in peak power as a function of the number of customers per transformer, at 100\% EV fleet share. For the median transformer in an LV network, the EVs raise its annual peak by 44\% with \emph{Logged}, by 166\% with \emph{No tariff} and by 70\% with \emph{Power tariff}. Charging against the spot price alone more than doubles the peak in four in five LV networks (82\%), since the EVs charge in the same low-price hours. The power tariff more than halves this increase, but the increase remains about 1.6 times that of logged charging. The tariff lowers the peak of each household, but all households respond to the same prices, so their charging still coincides in the same hours, whereas logged charging is spread over the evening and night by differing habits. 

The dependence on the size of the LV network differs between the cases. With logged charging, the increase falls from 62\% in LV networks with 1--5 customers to 30\% in networks with 21--50 customers, as the charging of more households coincides less. In the optimized cases, it is instead largest in LV networks with 21--50 customers (217\% with \emph{No tariff}, 101\% with \emph{Power tariff}) and smaller in the smallest networks. Overall, the power tariff gives a lower peak than logged charging in a third of LV networks (14\% of customers): in 80\% of the networks with 1--5 customers, but in only about 1\% of those with 6--50.

Fig.~\ref{fig:increase}(b) shows the increase as a function of the share of apartments among the customers of the cell. The increase falls with the share of apartments, partly because apartment households have fewer EVs (lower panel). The three cases converge in cells dominated by apartment buildings, because EV charging in apartment buildings follows the logged profiles in all cases (Section~\ref{sec:ev}). 

\subsection{Violations at increasing EV fleet shares}\label{sec:res_sweep}

\begin{figure*}[pos=!tbp]
\centering
\includegraphics[width=\textwidth]{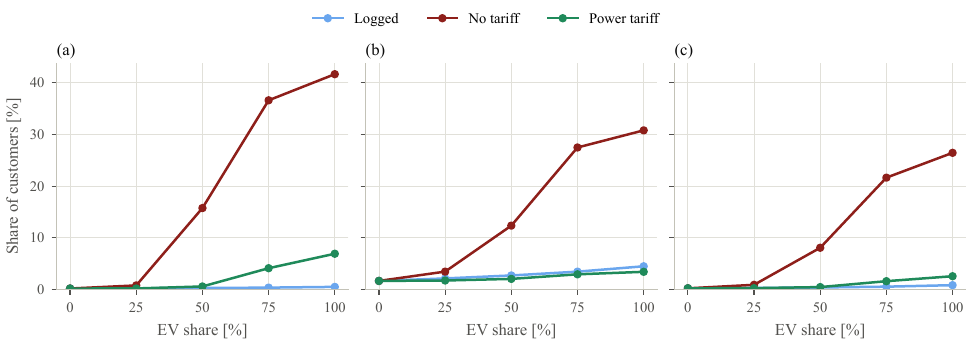}
\caption{Share of customers whose LV network has at least one violation during the year as a function of the EV fleet share, per charging case: (a) transformer overload, (b) undervoltage at the furthest customer and (c) cable overload.}
\label{fig:sweep}
\end{figure*}

Fig.~\ref{fig:sweep} shows the share of customers whose LV network has at least one violation during the year as a function of the EV fleet share, and Table~\ref{tab:main} summarizes the results at 100\%. 
With \emph{Logged} charging, violations increase gradually with the fleet share, to 0.5\% of customers connected to an LV network with transformer overload, 4.5\% with undervoltage and 0.8\% with cable overload at 100\% EV fleet share. Undervoltage is the most common violation. With \emph{No tariff}, the shares hardly differ from logged charging at 25\% EVs, but rise steeply above this level. At 50\%, 16\% of customers are in LV networks with transformer overload, and at 100\%, 42\%, together with 31\% with undervoltage and 26\% with cable overload. The steep rise reflects a threshold: once enough houses in the same network have EVs that charge in the same low-price hours, the design margin of the transformer is exceeded in many networks at once. Above 75\%, the increase levels off, as most LV networks that serve houses already have violations. With the power tariff, thermal violations stay below 1\% of customers up to 50\% EVs, but diverge from logged charging between 50\% and 75\% EVs: transformer overload rises to 4.1\% at 75\% EVs and 6.9\% at 100\% EVs, and cable overload to 2.6\% at 100\% EVs. The power tariff thus reduces the share of customers with transformer overload by 83\% compared with \emph{No tariff} at 100\%, but leaves about 14 times as many customers affected as logged charging. For undervoltage, in contrast, the power tariff gives fewer violations than logged charging at every fleet share, with 3.4\% of customers at 100\%. 
Shares weighted by LV networks instead of customers are given in Appendix~\ref{app:sweep}.

Violations with \emph{No tariff} are also persistent: 42\% of customers are in LV networks with transformer overload in most draws, against 5.8\% with \emph{Power tariff}, whereas with logged charging most affected customers are in LV networks with rare violations (Appendix~\ref{app:prevalence}).

The undervoltage shares count every interval below 0.95~$U_{\mathrm{N}}$, regardless of how far the voltage falls. The lowest voltage at the furthest customer is below $0.95\,U_{\mathrm{N}}$ in the median draw for 2.8\% of customers with \emph{Logged}, 32\% with \emph{No tariff} and 2.1\% with \emph{Power tariff}, and in an unfavorable draw (the 5th percentile of the lowest voltage) for 13\%, 39\% and 11\%. 
It falls below $0.90\,U_{\mathrm{N}}$, the lower limit of EN~50160, for practically no customers in the median draw and for at most 0.1\% of customers in an unfavorable draw in any charging case (Appendix~\ref{app:loading}). For almost all customers, the lowest voltages stay above about 0.92--0.93~$U_{\mathrm{N}}$, because the feeders are designed for a voltage drop of at most 5\% at the design load and, with \emph{No tariff}, the coincident load per house saturates near the main fuse. The undervoltage reported here therefore mainly indicates that the design margin of the feeders is used up. Since the model assumes balanced three-phase loading and one representative feeder per LV network that meets the current design criteria, the share of customers below $0.90\,U_{\mathrm{N}}$ is likely underestimated.

\begin{table}[pos=!tbp]
\centering
\caption{Share of customers whose LV network has at least one violation during the year at 100\% EVs, and the median increase in peak power over LV networks.}
\label{tab:main}
\begin{tabular}{lccc}
\hline
 & Logged & No tariff & Power tariff \\
\hline
Transformer overload [\%] & 0.5 & 42 & 6.9 \\
Undervoltage [\%] & 4.5 & 31 & 3.4 \\
Cable overload [\%] & 0.8 & 26 & 2.6 \\
\hline
Increase in peak power [\%] & 44 & 166 & 70 \\
\hline
\end{tabular}
\end{table}

\subsection{Where violations occur}\label{sec:res_where}
 
\begin{figure*}[pos=!tbp]
\centering
\includegraphics[width=\textwidth]{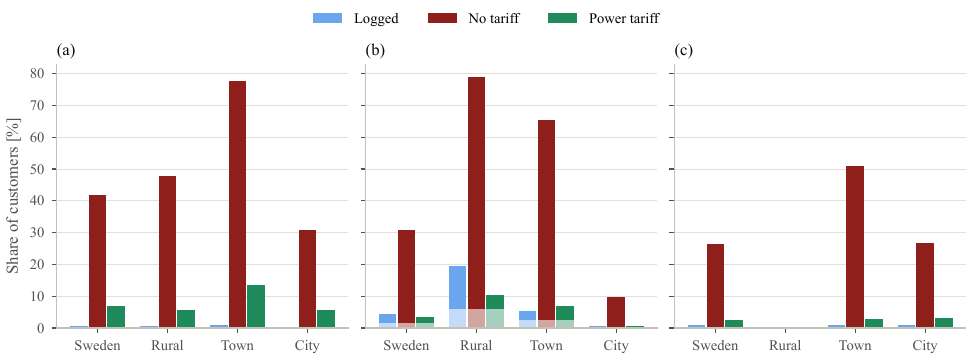}
\caption{Share of customers whose LV network has at least one violation during the year at 100\% EVs, per area type and charging case: (a) transformer overload, (b) undervoltage at the furthest customer, (c) cable overload. The lighter part of each bar occurs already without EVs.}
\label{fig:where}
\end{figure*}

Fig.~\ref{fig:where} shows the share of customers affected by violations per area type. The violation types occur in different areas. Undervoltage is concentrated in rural areas, where 6.0\% of customers are already in LV networks with undervoltage without EVs. With EVs, this share rises to 19\% with \emph{Logged}, 79\% with \emph{No tariff}, and 10\% with \emph{Power tariff}. Transformer overload is most frequent in towns in all three charging cases, with 0.8\%, 78\% and 14\% of customers in an affected network. Cable overload practically does not occur in rural areas, where the feeders reach the voltage limit before their thermal limit.

The size of the LV network determines how the power tariff compares with logged charging. In networks with up to 20 customers, the power tariff gives less undervoltage than logged charging, about half or less. In networks with 21--150 customers, it gives far more transformer overload, about 15 times as much (Appendix~\ref{app:where}). Since 87\% of customers live in networks with more than 20 customers, the power tariff gives more transformer overload than logged charging nationally. In LV networks with more than 150 customers, which consist mainly of apartment buildings, the cases hardly differ. For the same reason, cities show the smallest differences between the cases, although 31\% of city customers still have transformer overload with \emph{No tariff}.

Per municipality, the share of customers with transformer overload ranges from 0 to 3.1\% with logged charging, from 0.2 to 99\% with \emph{No tariff} and from 0 to 26\% with \emph{Power tariff} (Appendix~\ref{app:where}).
Differences between bidding zones are smaller than between area types (Appendix~\ref{app:where}). With \emph{No tariff}, the share of customers with transformer overload ranges from 38\% in SE3 to 50\% in SE4, and with \emph{Power tariff} from 5.6\% in SE2 to 8.9\% in SE4. 

\subsection{How often and when violations occur}\label{sec:res_when}
An LV network counts as affected after a single violating interval in a year, so the number of violating intervals shows how often violations recur. Averaged over all draws, a customer's LV network experiences 0.6 intervals with transformer overload per year with \emph{Logged}, 110 with \emph{No tariff}, and 5.4 with \emph{Power tariff}, and 3.9, 45, and 4.7 intervals with undervoltage, respectively (Appendix~\ref{app:severity}). With the power tariff, fewer customers are affected by undervoltage than with logged charging, but in the affected networks violations recur more often, so that the number of violating intervals per customer is somewhat higher, concentrated in the winter nights.

\begin{figure*}[pos=!tbp]
\centering
\includegraphics[width=\textwidth]{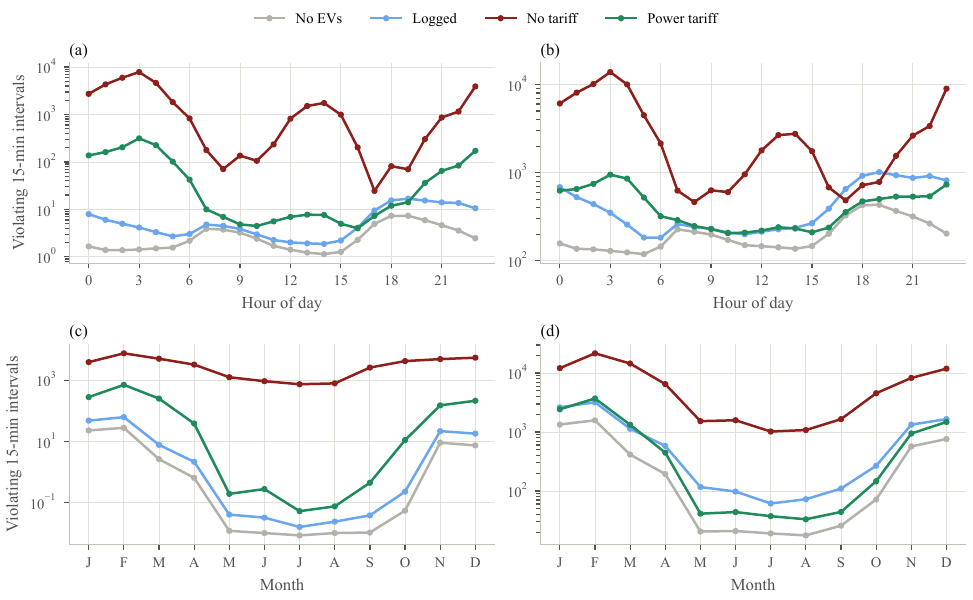}
\caption{Violating 15-min intervals per year at 100\% EVs, by hour of the day (a, b) and by month (c, d), for transformer overload (a, c) and undervoltage at the furthest customer (b, d), per charging case and without EVs. Values are the number of violating intervals in an average simulated year, summed over the representative LV network of each cell and scaled to 1,000 cells.}
\label{fig:when}
\end{figure*}

Fig.~\ref{fig:when} shows when transformer overload and undervoltage occur during the day and the year. Without EVs, violations occur at the household peaks, mainly between 17:00 and 20:00 and to a lesser extent in the morning, and almost only from November to March. Logged charging raises these evening peaks and extends them into the night, as charging tends to start in the evening and continues after midnight, while the seasonal pattern remains that of the household load: violations occur mainly in winter.

With \emph{No tariff}, the timing changes completely. Violations occur mainly between 22:00 and 05:00, with a maximum around 03:00, when the spot price is lowest, and a second maximum around 13:00--14:00, when low midday prices also attract charging. In the evening hours of the household peak, \emph{No tariff} gives fewer undervoltage intervals than logged charging, since charging is moved away from these high-price hours. Violations also occur throughout the year: in the summer months, transformer overloads with \emph{No tariff} are still more frequent than in the winter months with any other case. 

With \emph{Power tariff}, violations follow the same nightly pattern as \emph{No tariff}, with a maximum around 03:00, but with far fewer violating intervals. The midday maximum largely disappears. Unlike with \emph{No tariff}, the violations remain concentrated in winter, from November to March, when the charging in the night coincides with high heating demand. Compared with logged charging, the power tariff moves undervoltage from the evening into the night and reduces it in spring, summer and autumn, while in winter the two give a similar number of violating intervals.

\subsection{Transformer reinforcement}\label{sec:res_reinforcement}

\begin{figure}[pos=!tbp]
\centering
\includegraphics[width=\textwidth]{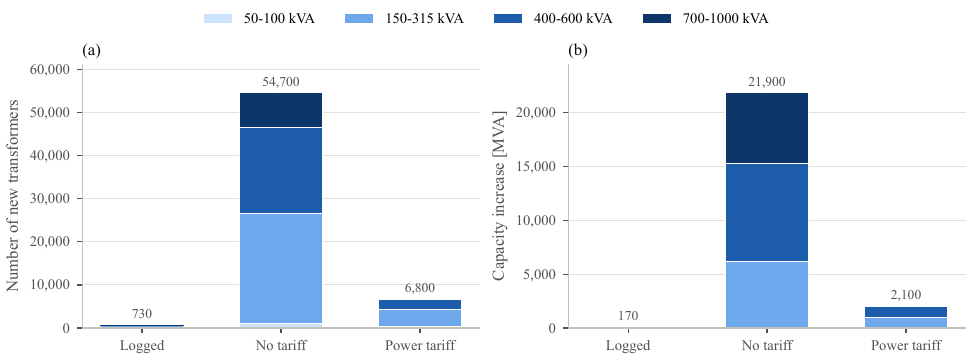}
\caption{Transformer reinforcement at 100\% EVs per charging case: (a) number of new transformers (added or replacing an existing one) and (b) capacity increase (added capacity plus net increase of replacements), per size class of the new transformer.}
\label{fig:reinforcement}
\end{figure}

Fig.~\ref{fig:reinforcement} shows the transformer reinforcement needed at 100\% EVs; all values are given in Table~\ref{tab:reinforcement} (Appendix~\ref{app:reinforcement}). With logged charging, 730 LV networks (0.5\%) need reinforcement, with a total capacity increase of 170~MVA, or 0.6\% of the installed transformer capacity. With \emph{No tariff}, 53,900 LV networks, about a third of all, need reinforcement, and the capacity increase of 21.9~GVA corresponds to 72\% of the installed capacity. The power tariff reduces this to 6,800 LV networks and 2.1~GVA (7\% of the installed capacity), i.e., it avoids about 90\% of the reinforcement needed with \emph{No tariff}.

The type of reinforcement reflects where the overloads occur. With logged charging and with \emph{No tariff}, about 30\% of the cells needing reinforcement have a single transformer that is replaced by a larger one, whereas with the power tariff this is the case for only 2.9\% of the cells, and the reinforcement consists almost entirely of added transformers. This agrees with the results of Section~\ref{sec:res_where}: with the power tariff, the overloads are concentrated in the larger LV networks, in cells with several transformers, whereas with logged charging and \emph{No tariff} they also occur in the small networks with only a transformer per cell. Most new transformers are 150--315~kVA in all cases (47--62\% of the new transformers). With \emph{No tariff}, larger units are also needed, and 32\% of the added capacity is transformers of 700--1,000~kVA.

The logged case represents today's charging behavior scaled to 100\% EVs, and \emph{No tariff} represents a full response of all house EVs to the spot price; the two thus bracket the reinforcement need for price-responsive charging without a power tariff, and the power tariff case lies between them.

\section{Discussion}\label{sec:discussion}
 
\subsection{Interpretation of the results}

The results show two effects of price-responsive charging. Charging against the spot price alone concentrates the charging of all EVs in the same low-price hours, which more than doubles the peak power of the median transformer in an LV network and leads to violations in LV networks serving 42\% of customers at 100\% EVs. This agrees with the high peaks found for cost-minimized charging in~\cite{Veldman2015, Zaferanlouei2022}: in this study violations with \emph{No tariff} rise steeply above 25\% EVs, and in~\cite{Zaferanlouei2022}, cost-minimized charging caused thermal violations above 36\% EVs in a Norwegian grid.

Adding a power tariff lowers each household's own peak, but all households still respond to the same prices, so their charging still coincides in the same nighttime hours. The peak of the median LV network is therefore 36\% lower than with \emph{No tariff}, but in LV networks with 6--150 customers, which serve two thirds of customers, the increase is 2 to 3.5 times that with logged charging, whose timing is spread by differing habits. This confirms at national scale what was anticipated in~\cite{Sandstrm2023}: individual peaks become lower, but the aggregated load in low-price hours can still be high. It also illustrates the concern of~\cite{ElGohary2023} that an individual peak is a weak proxy for the loading of the local grid.

How well the individual price signal works depends on the size of the LV network. Where a transformer serves a few customers, the peak of one household largely sets the peak of the network, and the power tariff gives fewer violations than logged charging, in particular less undervoltage. Where it serves tens to hundreds of customers, the coincident peak of several households creates the network peak, and the power tariff results in more than ten times as many customers with transformer overload as logged charging. A study of a single small feeder would thus find the power tariff more beneficial than a study of a single larger network, which illustrates why the regional heterogeneity found in~\cite{Schachler2021} calls for national-scale assessments.
 
\subsection{Level of the power tariff}
The power tariff was set to 7.4~€/kW per month, the average of 24 Swedish DSOs. A separate grid simulation for other tariff levels was not performed, because the charging behavior that the tariff induces is insensitive to its level. In the sensitivity analysis of~\cite[App.~A5.1]{Lundblad2026preprint}, lowering the tariff to the lowest level among the DSOs, 1.7~€/kW per month, gave the same household peaks and the same ranking of the tariff designs. The reason is that the savings resulting from following the spot price are small compared with charging that also keeps the household peak low~\cite{Lundblad2026preprint}. Once a power tariff is present, avoiding a new peak therefore dominates the charging decision, and the spot price only determines the timing within the remaining margin. The grid impacts of the power tariff case studied here can therefore be expected to hold for the range of tariff levels currently applied in Sweden. A higher tariff level would also not be expected to remove the remaining violations, since these result from the coincidence of many households' charging rather than from high individual peaks. This conclusion assumes that households respond to the tariff as a cost-minimizing optimization would, for example through a smart charger, which requires no active effort once installed.

\subsection{Implications for tariff design and DSOs} Swedish DSOs are subject to a revenue cap set by Ei, so a power tariff does not increase the total network revenue but redistributes it between customers~\cite{ellagen1997}. Customers who keep their peak low pay less, and customers with high or inflexible peaks pay more~\cite{Saele2017}. Whether this reallocation is fair depends on how closely an individual peak reflects the cost that the customer causes in the grid. As noted in~\cite{ElGohary2023}, individual peaks often occur at other times than the peak of the local grid, so a tariff on the individual peak also penalizes peaks that do not stress the grid. The results of this study support this concern for EV charging. In LV networks without overload, the tariff limits individual peaks even where the grid could accommodate them, which restricts the flexibility of EV charging and prevents it from using hours with low spot prices, although the power system has spare capacity. 

A power tariff is effective at limiting the damage of spot-price charging, and in this sense it could be preferable to no tariff at all: at 100\% EVs, it avoids about 90\% of the transformer reinforcement needed with \emph{No tariff}. As long as households optimize their charging against a common price, however, a tariff on the individual peak cannot prevent their charging from coinciding. Limiting the loading of larger LV networks requires a signal that reflects the coincident load, for example a tariff on the combined peak of a group of households~\cite{Lundblad2026preprint}, or a tariff or capacity limit that varies with the loading of the local grid~\cite{Gan2013, Daneshzand2023}. Since the coincidence grows with the share of EVs that respond to prices, the issue becomes increasingly important as smart charging spreads: in this study, the thermal violations with the power tariff diverge from those of logged charging between 50\% and 75\% EVs, assuming all EVs in houses optimize their charging. DSOs that plan grid reinforcement based on today's charging patterns may therefore underestimate the needs, since the charging of today's EV owners, represented by the logged case, has a much lower coincidence than cost-optimized charging.

Signals that limit charging only where and when the local grid is congested would let EVs follow the spot price elsewhere. Whether it is cheaper to reinforce LV grids so that all EVs can charge in the lowest-price hours, or to limit the coincidence of their charging, depends on the cost of reinforcement compared with the system value of the shifted energy, which was not quantified in this study.

The results from this study show that optimizing EV charging under a household power tariff can substantially reduce the peak power in an LV network. However, since controlling EV charging alone is sufficient to achieve most of this reduction, an alternative to household-level power tariffs may be a tariff that specifically targets flexible loads. This could avoid penalizing customers whose demand is largely inflexible and who may face higher tariff costs while lacking the financial means to invest in flexible technologies.
 
\subsection{Transferability to other countries}
The quantitative results are specific to Swedish conditions: three-phase connections with main fuses of 16--25~A, a large share of electric heating, mostly underground cables, and the Swedish design rules used to dimension the synthetic grids. The method itself can be applied elsewhere. REGAL and MANORS are open source~\cite{Lundblad2024b, ChalmersEnergySystems2026}, and applying them to another country requires data on population, dwelling types and car ownership; measured household load profiles with their fuse sizes; logged EV driving and charging; spot prices and network tariffs; and the national design rules and component catalogues for LV grids, which are needed to recalibrate the grid against data from local DSOs~\cite{Lundblad2024a}. In countries with single-phase connections or overhead lines, the balance between thermal and voltage violations is likely to differ. The dependence on the size of the LV network found here is likely to transfer, since it follows from the coincidence of households responding to a common price rather than from Swedish design rules.
 
\subsection{Extensions and future work}
Technologies whose operation is optimized together with the household can be added with
few changes to the grid model. Future work will extend MANORS to optimize EV charging together with rooftop PV and battery storage, both for individual houses and for apartment buildings. Heat pump flexibility could be represented in MANORS by treating the heat pump's operation as a decision variable alongside EV charging. Future work will also add commercial customers, including public charging points, to the synthetic grids. The design rules take only load as input, namely the number of customers, the energy per customer, and a design margin, so commercial customers can enter in the same way as residential ones: they are included when the number of transformers is determined, their energy is added to the dimensioning load, and their profiles are added to the simulation. The rules would then build stronger grids in mixed cells, and the validation could be extended from mainly residential cells to all cells in the DSO data. This will make it possible to evaluate a daytime tariff (07:00–20:00) on cells with both residential and non-residential load, where shifting charging to the night may provide greater benefits than in residential cells.
 
\subsection{Limitations}\label{sec:limitations}
\paragraph{Charging behavior} The optimized cases assume that every EV owner in a house charges with full foresight of prices, driving, and household demand, and that all of them respond to the price signals. Real smart chargers differ in their settings, and not all owners use them, which would spread the charging and reduce the coincidence. The optimized cases thus mark the upper end of the coincidence that a common price can create, and the logged case its lower end. Conversely, the logged case already contains some price response: 51 of the 139 owners have an hourly contract and charge more at night than the others; the coincidence of the logged charging is nevertheless much lower than in the optimized cases. Finally, EV charging does not affect the spot price in this study, but with very high EV shares, coordinated charging in low-price hours could raise these prices and spread the charging somewhat.

The spot prices of 2024 are hourly while the model uses 15-min timesteps, so the four quarter-hours of an hour have identical prices, and the solver decides how charging is distributed among them. Using spot prices at a 15-minute resolution in the optimization may increase the coincidence of the optimized charging and would align with the day-ahead market, which clears at 15-min resolution as of October 2025.
 
\paragraph{Input data} All household profiles were measured in southwestern Sweden, so the household load is likely underestimated in the colder northern parts of the country. The 139 logged EVs are reused across many households with different week shifts, and their behavior is also used for apartment dwellers. The logged apartment owners charge less at home than the house owners (1,700 against 2,500~kWh per year) and less at night. Using all owners therefore probably overestimates the home charging of EVs in apartment buildings in the logged case.

\paragraph{Apartment buildings} Some charging operators already steer the charging toward lower-price hours and limit the peak power of the building through load management, which is not represented. MANORS does not yet optimize the EVs of a building jointly against a building-level tariff, which would be needed to assess the tariffs for apartment buildings.
Furthermore, since not every residential building is equipped with chargers, apartment dwellers use public chargers to a larger extent, which often charge at higher power, typically 11--22~kW for AC chargers along streets and in parking areas~\cite{Kobayashi2025}. Where these are connected to the residential LV grid, mainly in cities, the peaks of apartment areas are therefore likely underestimated. 
 
\paragraph{Grid representation} Only residential customers are simulated, and the grid is dimensioned for the residential load only. In mixed areas, the modeled grid is therefore a residential-equivalent grid, with lower capacity than the real one, which also carries commercial loads such as shops, schools and offices. These loads often peak during daytime and may change both the level and the timing of the loading. The results for mixed areas thus describe how the EV charging of the residents loads a grid sized for those residents, not the real grid. Public charging is not represented in the grid: the energy charged away from home is not added to any LV grid, likely underestimating the loading of city grids.

Each cell is represented by one representative LV network, voltage is evaluated only at its furthest customer, and transformer and cable loading are computed from active power. The voltage at the transformer is fixed at the nominal value, whereas in practice the tap setting and activity in the medium-voltage grid shift it up or down. The grid is not reinforced as EVs are added, and the reinforcement analysis covers transformers only, not cables or measures against undervoltage. Thermal limits are evaluated against the rated capacity regardless of ambient temperature. 

\section{Conclusions}\label{sec:conclusions}

This paper assessed how optimizing EV charging under a household power tariff and spot prices affects thermal and voltage violations in residential LV grids in Sweden. EV charging optimized together with measured household loads was placed on synthetic grids with 5.1 million residential customers, calibrated against DSO data, and compared with real-world charging for EV fleet shares up to 100\%.

Charging against the spot price alone gives the largest increase in peak power and the most violations. At 100\% EVs, it more than doubles the peak power of the median transformer in an LV network, and 42\% of customers are connected to LV networks with at least one transformer overload during the year, against 0.5\% with logged charging. Because all EVs respond to the same prices, charging concentrates at night and on low-price days throughout the year, and the transformer capacity would have to increase by 72\% to accommodate it.

A power tariff on the monthly household peak removes most of these violations: it reduces the share of customers with transformer overload to 6.9\% and avoids about 90\% of the transformer reinforcement needed without a tariff. The tariff lowers each household's own peak, but their charging still coincides. As a result, the power tariff requires new transformer capacity equal to 7\% of the installed capacity (2.1~GVA), against 0.6\% with logged charging.

The effect of the power tariff depends on the size of the LV network. In networks with few customers, it gives less violations than logged charging. In networks with 21--150 customers, where coincident charging dominates the peak load, the share of customers with transformer overload is instead about 15 times as high as with logged charging. Results from a single network can thus not be generalized to a whole country. 

For regulators and DSOs, the results indicate that a power tariff on the individual peak is far better than no tariff when many EV owners optimize their charging against the spot price, but that it cannot by itself keep this charging within the capacity of existing LV grids at high EV shares. Tariffs that act on the aggregated peak of a group of households, or price signals that reflect the loading of local grid, are needed for that. Future work will evaluate such collective tariffs, optimize the EV charging in apartment buildings, add rooftop PV and battery storage, and add commercial customers to the synthetic grids.
 
\appendix
 
\section{Supplementary results}

\subsection{Grid design and calibration}\label{app:design}
REGAL dimensions the grid of each cell from its numbers of houses and apartments. With respect to~\cite{Lundblad2024a}, the free parameters were recalibrated in this work against the grid data described in Section~\ref{sec:validation}. The number of transformers in a cell with $C$ customers is $\mathrm{round}(\sqrt{C}/k)$ with $k = 3.0$, i.e., a transformer serves about $3\sqrt{C}$ customers, which best reproduces the number of transformers in the mainly residential cells of the calibration data. The required transformer capacity is the coincident peak of its customers from Velander's formula, $P = k_1 W + k_2 \sqrt{W}$, with $W$ the annual energy of the customers, multiplied by a design margin. The design energy is 18.6~MWh per house, corresponding to an electrically heated house, and 3.55~MWh per apartment, including the common-area consumption~\cite{Zimmermann2009}. The margin is 1.7 in cells with up to 20 customers, which always have a single transformer, 1.2 in larger cells and 1.5 in cells where apartments make up at least 80\% of the customers; these margins were set so that the simulated loading of the transformers matches the measured loading of the DSO's transformers (Table~\ref{tab:calibration}). The transformer is the smallest standard size that exceeds the required capacity; the resulting grids contain sizes between 50 and 1000~kVA. Cables are chosen as the thinnest combination that meets the voltage drop at the design load (at most 5\% at the furthest customer), the earth-loop impedance limit for fuse tripping and the thermal limit for the design current, and a margin of 1.1. The thinnest cable used is Al50, since thinner cables are rare in the calibration data.

\paragraph{Robustness of the calibration} To test how much the design parameters depend on the data of particular regions, each of the seven regions of the calibration data is left out in turn, the parameters are refitted on the other regions and compared with the values used (Table~\ref{tab:loro}). The constant $k$ is refitted so that the number of transformers is correct on average, the transformer margins are refitted to the median and 90th percentile of the measured transformer loading, and the cable margin to the length distribution of cable types. The refitted values differ from the ones used by at most 0.2. With $k = 3.0$, the per-cell error in the number of transformers in the left-out region changes by at most one percentage point. Fitted on a single region, the values differ more: the most urban region alone would give $k = 3.5$, i.e., fewer and larger transformers. The calibrated rules thus represent an average design practice rather than that of any region.

\paragraph{Transformer loading and consumption factor} The transformer margins and the standard deviation $\sigma$ of the cell consumption factor are calibrated against the measured loading of the DSO's transformers. For each secondary substation, the data give the third-highest load recorded 2022--2025 relative to the rated power. The calibration is performed on cells where residential customers account for more than 80\% of the metered energy. 

The measured values are compared with the simulated annual peak without EVs in an unfavorable draw (95th percentile over the draws). As a real area keeps its consumption factor from year to year, the fixed-level runs are used: the unfavorable-draw loading of each level is weighted by the probability of the level. REGAL then gives a median loading of 0.38, a 90th percentile of 0.63 and 2.6\% of transformers loaded above 0.8, against 0.36, 0.65 and 3.0\% in the data (Table~\ref{tab:calibration}).

\paragraph{Residential energy use} The annual residential load resulting from the model is validated against the metered annual consumption of permanent dwellings in the calibration data. Holiday homes are excluded, since REGAL models only permanent dwellings. The comparison is made at municipality level, in the 38 municipalities in which the data cover at least 80\% of the cells. The ratio of simulated to metered energy has a median of 1.05, and 34 of the 38 municipalities lie within $\pm 20\%$. Over all municipalities, the simulated energy is 14\% higher, mainly because of the largest municipality (1.02 without it). A likely reason is that the common-area electricity of apartment buildings is often registered as non-residential: counting property-management consumption in apartment-dominated cells as residential gives 1.00 for this municipality. Ratios below 0.9 occur in suburbs with large electrically heated houses and in coastal municipalities with many second homes. 

\begin{table}[pos=!htbp]
\caption{Calibration of the grid design against DSO data in mainly residential cells. Transformer loading: annual peak relative to the rated power.}
\label{tab:calibration}
\centering
\footnotesize
\begin{tabular}{lcc}
\hline
 & Model & Data \\
\hline
\multicolumn{3}{l}{\emph{House areas (residential share of energy $> 0.8$)}} \\
Transformer loading, median & 0.38 & 0.36 \\
Transformer loading, 90th percentile & 0.63 & 0.65 \\
Share of transformers loaded above 0.8 [\%] & 2.6 & 3.0 \\
Spread of loading (90th percentile / median) & 1.65 & 1.78 \\
Share of cable length Al150 or thicker [\%] & 40 & 29 \\
\quad cells with more than 50 customers [\%] & 65 & 68 \\
\multicolumn{3}{l}{\emph{Apartment areas (residential share of energy $> 0.6$)}} \\
Spread of loading (90th percentile / median) & 1.40 & 1.67 \\
\hline
\end{tabular}
\end{table}

\begin{table}[pos=!htbp]
\caption{Design parameters fitted on all seven regions of the calibration data, the range of the values refitted with each region left out in turn, and on each region alone.}
\label{tab:loro}
\centering
\footnotesize
\begin{tabular}{lccc}
\hline
Parameter & Used & One region left out & Single regions \\
\hline
Customers per transformer, $k$ & 3.0 & 2.8--3.1 & 2.5--3.5 \\
Transformer margin, $\le 20$ customers & 1.7 & 1.5--1.6 & 1.5--1.8 \\
Transformer margin, $> 20$ customers & 1.2  & 1.2--1.25 & 1.1--1.25 \\
Cable margin & 1.1 & 1.0--1.1 & 1.0--1.2 \\
\hline
\end{tabular}
\end{table}

\subsection{Levels of the consumption factor in the fixed-level runs}\label{app:levels}
Table~\ref{tab:levels} gives the levels used in the fixed-level runs (Section~\ref{sec:metrics}). The three highest levels together have a probability of 0.3\% but are included because the reinforcement need is concentrated in high-consumption cells.

\begin{table}[pos=!htbp]
\caption{Levels of the consumption factor in the fixed-level runs: interval of the standard normal variable $z_c$, probability (weight) and mean consumption factor of houses and apartment buildings. Each level is simulated with 112 household draws per cell.}
\label{tab:levels}
\centering
\begin{tabular}{clccc}
\hline
Level & $z_c$ interval & Weight & Houses & Apartments \\
\hline
1 & $< -0.5$ & 0.30854 & 0.687 & 0.784 \\
2 & $-0.5$ to $0.5$ & 0.38292 & 0.959 & 0.982 \\
3 & $0.5$ to $1.25$ & 0.20289 & 1.231 & 1.159 \\
4 & $1.25$ to $1.75$ & 0.06559 & 1.487 & 1.316 \\
5 & $1.75$ to $2.25$ & 0.02783 & 1.722 & 1.451 \\
6 & $2.25$ to $2.75$ & 0.00924 & 1.995 & 1.600 \\
7 & $2.75$ to $3.25$ & 0.00240 & 2.312 & 1.765 \\
8 & $3.25$ to $3.75$ & 0.00049 & 2.678 & 1.948 \\
9 & $> 3.75$ & 0.00009 & 3.170 & 2.178 \\
\hline
\end{tabular}
\end{table}

\subsection{Convergence and seed dependence}\label{app:convergence}
The draws of each cell are generated from random streams seeded per cell, so the results are reproducible. In the random-level runs, repeating a run with a different seed changes the shares of customers with transformer overload, undervoltage or cable overload by less than 1.2\%, and increasing the number of draws per cell from 100 to 1,000 changes them by less than 1.5\%. The median and unfavorable-draw peak loadings change by less than 2\% in both tests. These differences are far below the differences between the charging cases. The fixed-level runs, with 1,008 draws per cell, are more stable still: between seeds, their shares of affected customers differ by less than 0.5\%, and the number of transformers whose unfavorable-draw loading exceeds their rating differs by 0.3\%.

\subsection{Share of customers and LV networks with violations}\label{app:sweep}
Table~\ref{tab:sweep} gives the shares of LV networks affected by violations next to the share of customers connected to an LV network with violations. Undervoltage is more frequent when weighted by LV networks than by customers, since it occurs mainly in small LV networks with few customers: at 100\% EVs, 15\% of LV networks have undervoltage with logged charging, 58\% with \emph{No tariff} and 7.1\% with \emph{Power tariff}. Thermal violations, which occur mainly in larger LV networks, are instead less frequent per LV network than per customer.

\begin{table}[pos=!htbp]
\caption{Shares of customers and of LV networks with at least one violation during the year, per charging case and EV fleet share. T: transformer overload, U: undervoltage, C: cable overload.}
\label{tab:sweep}
\centering
\footnotesize
\setlength{\tabcolsep}{3.5pt}
\begin{tabular}{l r *{6}{wc{1.8em}}}
\hline
 & EV share & \multicolumn{3}{c}{Customers [\%]} & \multicolumn{3}{c}{LV networks [\%]} \\
Case & [\%] & {T} & {U} & {C} & {T} & {U} & {C} \\
\hline
No EVs & 0 & 0.2 & 1.7 & 0.2 & 0.2 & 4.3 & 0.1 \\
\hline
Logged & 25 & 0.2 & 2.1 & 0.3 & 0.2 & 6.1 & 0.1 \\
 & 50 & 0.3 & 2.7 & 0.4 & 0.2 & 8.2 & 0.2 \\
 & 75 & 0.4 & 3.5 & 0.5 & 0.3 & 11 & 0.2 \\
 & 100 & 0.5 & 4.5 & 0.8 & 0.4 & 15 & 0.3 \\
\hline
No tariff & 25 & 0.8 & 3.5 & 0.9 & 0.9 & 10 & 0.4 \\
 & 50 & 16 & 12 & 8.1 & 14 & 29 & 4.3 \\
 & 75 & 37 & 27 & 22 & 32 & 54 & 12 \\
 & 100 & 42 & 31 & 26 & 35 & 58 & 14 \\
\hline
Power & 25 & 0.2 & 1.7 & 0.3 & 0.2 & 4.5 & 0.1 \\
tariff & 50 & 0.6 & 2.1 & 0.5 & 0.5 & 5.2 & 0.2 \\
 & 75 & 4.1 & 2.9 & 1.6 & 3.3 & 6.4 & 0.8 \\
 & 100 & 6.9 & 3.4 & 2.6 & 5.0 & 7.1 & 1.2 \\
\hline
\end{tabular}
\end{table}

\subsection{Persistence of violations}\label{app:prevalence}

\begin{figure*}[pos=!tbp]
\centering
\includegraphics[width=\textwidth]{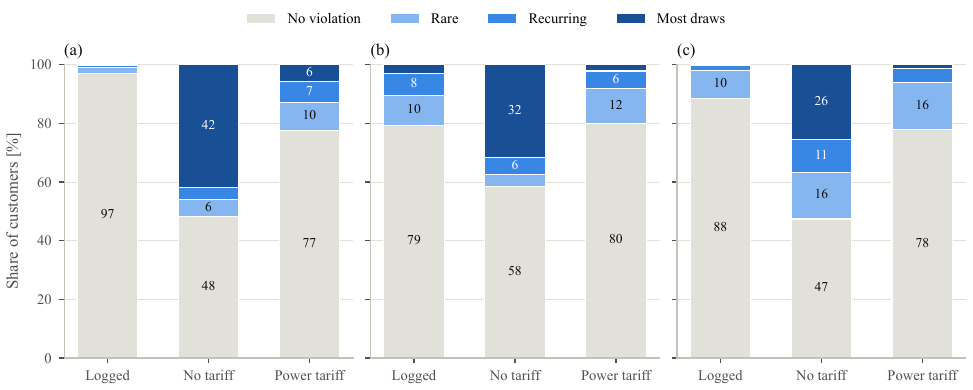}
\caption{Share of customers at 100\% EVs by how often their LV network has a violation: no violation ($p_{c,x,b} = 0$), rare ($0 < p_{c,x,b} < 0.1$), recurring ($0.1 \leq p_{c,x,b} < 0.5$) and in most draws ($p_{c,x,b} \geq 0.5$), for (a) transformer overload, (b) undervoltage at the furthest customer and (c) cable overload.}
\label{fig:prevalence}
\end{figure*}

To show how persistent violations are across draws, LV networks are grouped by how often they experience violations. In the fixed-level runs, $p_{c,x,b}$ is classified separately at each level $b$: never, $p_{c,x,b} = 0$; rare, $0 < p_{c,x,b} < 0.1$; recurring, $0.1 \leq p_{c,x,b} < 0.5$; and in most draws, $p_{c,x,b} \geq 0.5$. The share of customers in class $k$ is
\begin{equation}
\frac{\sum_c \sum_b w_b\, C_c\, \mathds{1}\left[\mathrm{class}(p_{c,x,b}) = k\right]}{\sum_c C_c}.
\end{equation}
Fig.~\ref{fig:prevalence} shows the result at 100\% EVs. With logged charging, most customers affected by transformer or cable overload are in LV networks with rare violations.

\subsection{Loading and voltage distributions}\label{app:loading}
\begin{figure*}[pos=!tbp]
\centering
\includegraphics[width=\textwidth]{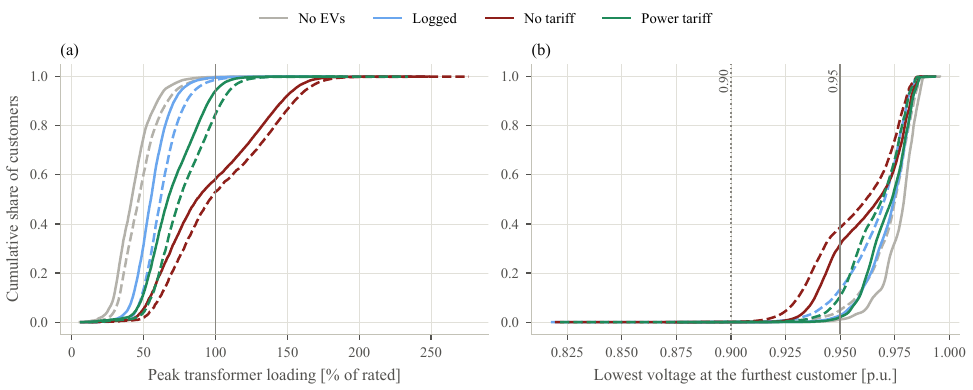}
\caption{Cumulative distributions over customers of (a) the peak transformer loading relative to the rating and (b) the lowest voltage at the furthest customer, per charging case and without EVs; median over household draws (solid) and 95th percentile of loading/ 5th of voltage (dashed).}
\label{fig:loading_voltage}
\end{figure*}

Fig.~\ref{fig:loading_voltage} shows the distributions over customers of the peak transformer loading and of the lowest voltage at the furthest customer in the median and unfavorable draw. In the median draw, 0.1\% of customers are in LV networks loaded above their rating without EVs, 0.3\% with logged charging, 42\% with \emph{No tariff} and 5.7\% with \emph{Power tariff}; in an unfavorable draw, 0.4\%, 1.3\%, 47\% and 15\%. The voltage at the furthest customer falls below $0.95\,U_{\mathrm{N}}$ mainly in rural areas: in an unfavorable draw, this is the case for 19\% of rural customers without EVs, 61\% with logged charging, 95\% with \emph{No tariff} and 33\% with \emph{Power tariff}. Below $0.90\,U_{\mathrm{N}}$, the share is at most 0.4\% of rural customers, with \emph{No tariff} in an unfavorable draw.

\subsection{Results per network size, bidding zone and municipality}\label{app:where}
\begin{figure*}[pos=!tbp]
\centering
\includegraphics[width=\textwidth]{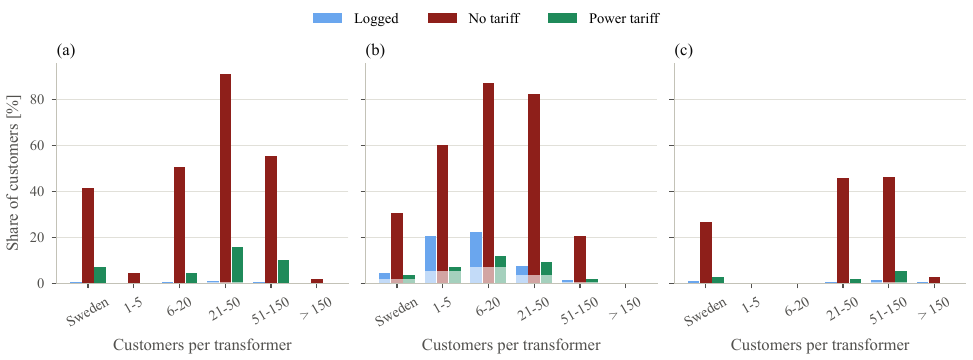}
\caption{Share of customers whose LV network has at least one violation during the year at 100\% EVs, per number of customers per transformer and charging case: (a) transformer overload, (b) undervoltage at the furthest customer and (c) cable overload.}
\label{fig:where_size}
\end{figure*}

Fig.~\ref{fig:where_size} shows the share of customers affected, by the size of their LV network, i.e. the number of customers per transformer. Undervoltage affects the largest share of customers in small LV networks, where the power tariff gives fewer violations than logged charging, whereas transformer and cable overload occur mainly in LV networks with 21--150 customers (Section~\ref{sec:res_where}). Practically no cable overload occurs in networks with up to 20 customers, and in networks with more than 150 customers, which consist mainly of apartment buildings, all cases affect less than 3\% of customers with each type of violation.

\begin{figure*}[pos=!tbp]
\centering
\includegraphics[width=\textwidth]{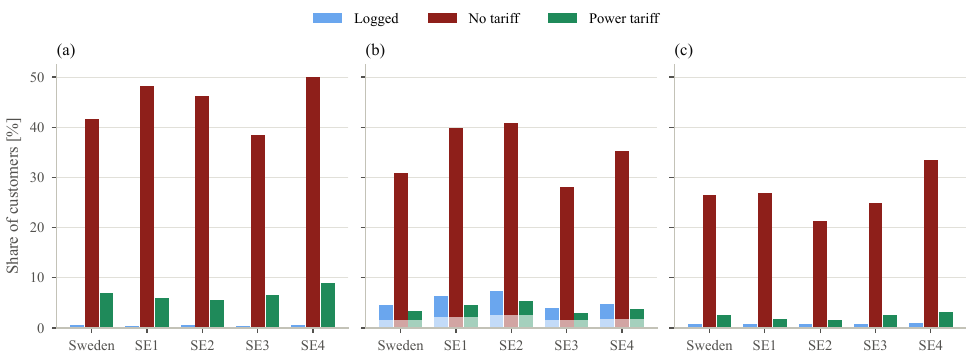}
\caption{Share of customers whose LV network has at least one violation during the year at 100\% EVs, per bidding zone and charging case: (a) transformer overload, (b) undervoltage at the furthest customer and (c) cable overload.}
\label{fig:where_area}
\end{figure*}

Fig.~\ref{fig:where_area} shows the results per bidding zone. SE3, which contains about 70\% of the population and of the apartments, has the lowest shares of undervoltage in all cases. In general, the results differ little between bidding zones.

\begin{figure*}[pos=!tbp]
\centering
\includegraphics[width=\textwidth]{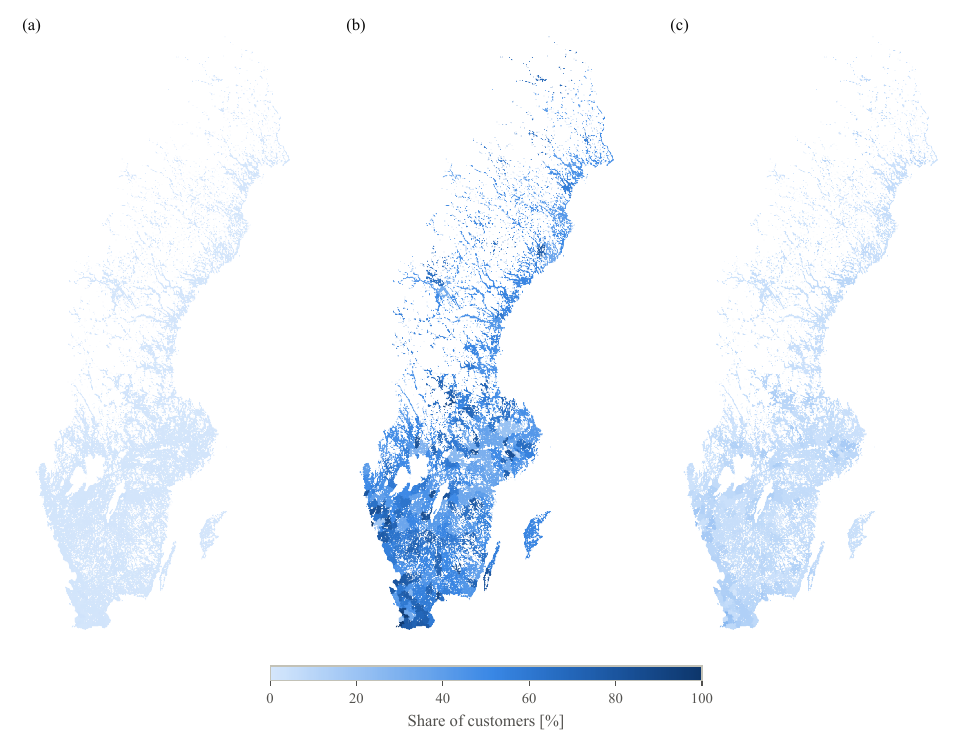}
\caption{Share of customers whose LV network has transformer overload in a year at 100\% EVs, for each charging case: (a) Logged, (b) No tariff, (c) Power tariff. Only populated cells, each in the color of its municipality.
}
\label{fig:map}
\end{figure*}

Fig.~\ref{fig:map} shows the share of customers with transformer overload per municipality. With \emph{No tariff}, the share is highest, above 85\%, in suburban municipalities dominated by houses, and lowest in the largest cities (11--21\%), where most customers live in apartment buildings.

\subsection{Severity of violations}\label{app:severity}
\begin{figure*}[pos=!htbp]
\centering
\includegraphics[width=\textwidth]{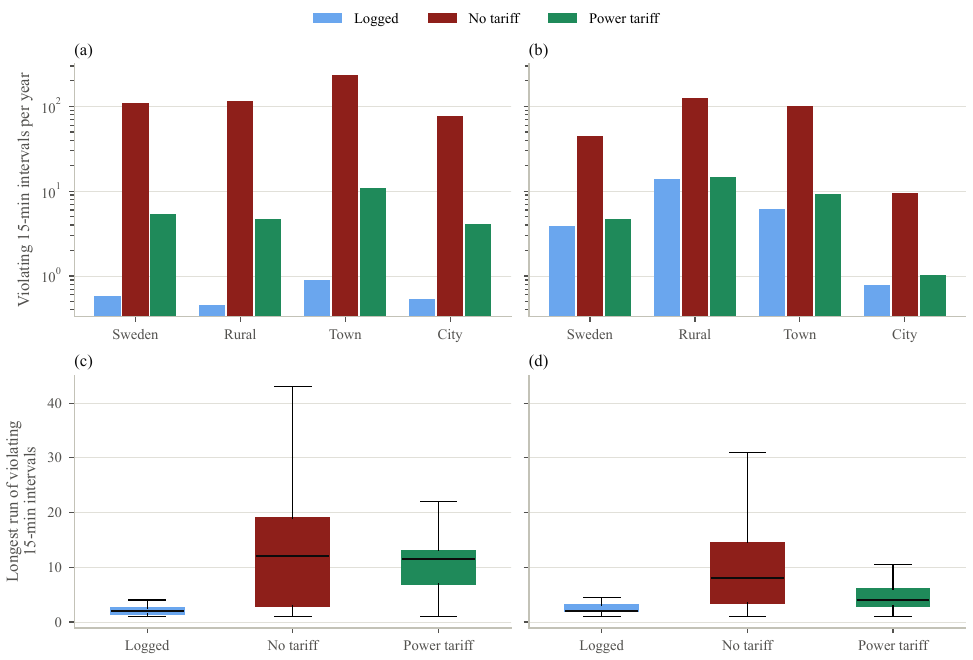}
\caption{Frequency and duration of violations at 100\% EVs, for transformer overload (a, c) and undervoltage at the furthest customer (b, d). (a, b) Number of violating 15-min intervals per year in a customer's LV network, averaged over all draws, nationally and per area type. (c, d) Longest run of consecutive violating intervals in the median violating draw of each LV network, for networks with a violation in at least 1 draw in 10.}
\label{fig:severity}
\end{figure*}

Fig.~\ref{fig:severity} shows how often and how long violations occur. With \emph{No tariff}, a customer's LV network has on average 230 intervals with transformer overload per year in towns and 77--114 in rural and city areas; with \emph{Power tariff}, 4--11, and with logged charging, below one. Undervoltage intervals are most frequent in rural areas, 14--15 per year with logged charging and \emph{Power tariff} and 126 with \emph{No tariff}. Where violations recur (at least 1 draw in 10), the longest uninterrupted transformer overload in the median violating draw lasts 2 intervals with logged charging, but about 12 intervals with both optimized cases: the power tariff confines transformer overloads to fewer LV networks and makes them less frequent, but not necessarily shorter, since some arise from charging that continues over several night hours. Undervoltage lasts 2 intervals with logged charging, 4 with \emph{Power tariff} and 8 with \emph{No tariff}.

\subsection{Transformer reinforcement}\label{app:reinforcement}
Table~\ref{tab:reinforcement} gives the transformer reinforcement at 100\% EVs following the criteria of Section~\ref{sec:reinforcement}.
Table~\ref{tab:reinforcement_sens} shows the transformer reinforcement for other reinforcement thresholds and target loadings. The number of LV networks to reinforce depends only on the threshold; the capacity increase depends on both. When the threshold or the target loading is varied, the capacity increase ranges from 80 to 390~MVA with logged charging, from 13 to 29~GVA with \emph{No tariff} and from 0.75 to 5.1~GVA with \emph{Power tariff}. The power tariff is more sensitive to the threshold than \emph{No tariff}, since its overloads are more often marginal: raising the threshold from 0 to 20\% removes 89\% of the LV networks to reinforce with \emph{Power tariff} and 34\% with \emph{No tariff}.

\begin{table}[pos=!htbp]
\centering
\caption{Transformer reinforcement at 100\% EVs (design load: 95th percentile of the annual peak; threshold 10\%; target loading 60\%).}
\label{tab:reinforcement}
\begin{tabular}{lccc}
\hline
 & Logged & No tariff & Power tariff \\
\hline
LV networks to reinforce  & 730 & 53,900 & 6,800 \\
\quad share of all [\%]  & 0.5 & 34 & 4.3 \\
Mean exceedance of rating [\%]  & 21 & 40 & 18 \\
Transformers added  & 640 & 48,100 & 6,700 \\
Transformers replaced  & 90 & 6,600 & 60 \\
Capacity increase [MVA]  & 170 & 21,900 & 2,100 \\
\quad share of installed [\%]  & 0.6 & 72 & 6.9 \\
\hline
\end{tabular}
\end{table}

\begin{table}[pos=!htbp]
\caption{Sensitivity of the transformer reinforcement at 100\% EVs to the reinforcement threshold (at a target loading of 60\%) and to the target loading (at a threshold of 10\%). Main setting in bold.}
\label{tab:reinforcement_sens}
\centering
\footnotesize
\setlength{\tabcolsep}{4pt}
\begin{tabular}{lrrr}
\hline
 & Logged & No tariff & Power tariff \\
\hline
\multicolumn{4}{l}{\emph{Capacity increase [MVA]}} \\
Threshold 0\%  & 390 & 23,900 & 5,100 \\
Threshold 5\%  & 260 & 22,800 & 3,500 \\
\textbf{Threshold 10\%, target 60\%}  & \textbf{170} & \textbf{21,900} & \textbf{2,100} \\
Threshold 20\%  & 80 & 19,700 & 750 \\
Target 50\%  & 240 & 29,000 & 2,900 \\
Target 70\%  & 130 & 17,100 & 1,600 \\
Target 80\%  & 100 & 13,200 & 1,100 \\
\hline
\multicolumn{4}{l}{\emph{LV networks to reinforce}} \\
Threshold 0\%  & 1,900 & 65,600 & 18,400 \\
Threshold 5\%  & 1,200 & 59,400 & 12,000 \\
\textbf{Threshold 10\%} & \textbf{730} & \textbf{53,900} & \textbf{6,800} \\
Threshold 20\%  & 290 & 43,600 & 2,000 \\
\hline
\end{tabular}
\end{table}

\section*{Acknowledgements}
The authors gratefully acknowledge the Norwegian Public Roads Administration, the Swedish Electromobility Centre, and the Swedish Energy Agency for financial support. They also thank several grid companies, including Mölndal Energi, for providing valuable input data and discussions.

\section*{Declaration of generative AI and AI-assisted technologies in the manuscript preparation process}
Claude (Anthropic) was used to write post-processing and plotting code and to improve the language and structure of the text. The authors reviewed and edited the content as needed and take full responsibility for the content of the article.

\bibliographystyle{elsarticle-num}
\bibliography{references}

\end{document}